\documentclass[a4paper,12pt]{article}
\usepackage{a4,amsmath,amssymb,axodraw,slashed,epsf}
\usepackage[square,comma,numbers,sort&compress]{natbib}
\usepackage[dvips]{color}

\begin{document}

%%%%%%%%%%%%%%%%%%%%  New Commands  %%%%%%%%%%%%%%%%%%%%%%%%%%%%%%%%%%%%%

\newcommand{\bed}{\begin{displaymath}}
\newcommand{\eed}{\end{displaymath}}
\newcommand{\beq}{\begin{equation}}
\newcommand{\eeq}{\end{equation}}
\newcommand{\bea}{\begin{eqnarray}}
\newcommand{\eea}{\end{eqnarray}}
\newcommand{\tgb}{{\rm tg}\beta}
\newcommand{\tga}{{\rm tg}\alpha}
\newcommand{\stgb}{{\rm tg}^2\beta}
\newcommand{\sia}{\sin\alpha}
\newcommand{\coa}{\cos\alpha}
\newcommand{\sib}{\sin\beta}
\newcommand{\cob}{\cos\beta}
\newcommand{\MS}{\overline{\rm MS}}
\newcommand{\st}{\tilde{t}}
\newcommand{\sgl}{\tilde{g}}
\newcommand{\Dmb}{\ensuremath{\Delta_b}}

\newcommand{\tb}{{\rm tg}\beta}
\newcommand{\gluino}{\widetilde{g}}
\newcommand{\squark}{\tilde{q}}
\newcommand{\Mq}[1]{m_{\tilde{q}_{#1}}}
\newcommand{\mq}{m_{q}}
\newcommand{\Aq}{A_q}
\newcommand{\mix}{\widetilde{\theta}}
\newcommand{\Lag}{\mathcal{L}}
\newcommand{\ls}{\lambda_{s}}
\newcommand{\lb}{\lambda_{b}}
\newcommand{\lt}{\lambda_t}
\newcommand{\Deltamb}{\Delta_b}
\newcommand{\DeltambQCD}{\Delta_b^{QCD}}
\newcommand{\DeltambELW}{\Delta_b^{elw}}
\newcommand{\mb}{m_{b}}
\newcommand{\mt}{m_{t}}
\newcommand{\bR}{b_R}
\newcommand{\bL}{b_L}
\newcommand{\sR}{s_R}
\newcommand{\sL}{s_L}
\newcommand{\At}{A_t}
\newcommand{\Sb}{\Sigma_b}
\newcommand{\Ao}[1]{A_0(#1)}
\newcommand{\Co}[5]{C_0(#1,#2;#3,#4,#5)}
\newcommand{\as}{\alpha_s}
\newcommand{\Mg}{m_{\tilde{g}}}
\newcommand{\Mb}[1]{m_{\tilde{b}_{#1}}}
\newcommand{\Mt}[1]{m_{\tilde{t}_{#1}}}
\newcommand{\NL}{\nonumber\\}
\newcommand{\bs}{\widetilde{b}}
\newcommand{\str}{\widetilde{s}}
\newcommand{\bsL}{\bs_L}
\newcommand{\bsR}{\bs_R}
\newcommand{\strL}{\str_L}
\newcommand{\strR}{\str_R}
\newcommand{\ts}{\widetilde{t}}
\newcommand{\tsR}{\widetilde{t}_R}
\newcommand{\tsL}{\widetilde{t}_L}
\newcommand{\higgsino}{\widetilde{h}}
\newcommand{\stopx}{\tilde{t}}
\newcommand{\sbottom}{\tilde{b}}
\newcommand{\strange}{\tilde{s}}
\newcommand{\gh}{g_b^h}
\newcommand{\gH}{g_b^H}
\newcommand{\gA}{g_b^A}
\newcommand{\ght}{\tilde{g}_b^h}
\newcommand{\gHt}{\tilde{g}_b^H}
\newcommand{\gAt}{\tilde{g}_b^A}
\newcommand{\stb}{{\rm tg}^2\beta}
\newcommand{\Order}[1]{{\cal{O}}\left(#1\right)}
\newcommand{\Msusy}{M_{SUSY}}
\newcommand{\gluon}{g}
\newcommand{\sttop}{\widetilde{t}}
\newcommand{\CF}{C_F}
\newcommand{\CA}{C_A}
\newcommand{\TR}{T_R}
\newcommand{\Tr}[1]{{\rm Tr}\left[#1\right]}
\newcommand{\dk}{\frac{d^nk}{(2\pi)^n}}
\newcommand{\dqq}{\frac{d^nq}{(2\pi)^n}}
\newcommand{\partderiv}[1]{\frac{\partial}{\partial#1}}
\newcommand{\kmu}{k_{\mu}}
\newcommand{\qmu}{q_{\mu}}
\newcommand{\eUV}{\epsilon}
\newcommand{\dMq}[1]{\delta\Mq{#1}}
\newcommand{\gs}{g_s}
\newcommand{\MSbar}{\overline{\rm MS}}
\newcommand{\gPhit}{\tilde{g}_b^{\Phi}}
\newcommand{\muR}{\mu_{R}}
\newcommand{\bbbar}{b\bar{b}}
\newcommand{\MPhi}{M_{\Phi}}
\newcommand{\dlt}{\delta\lambda_t}
\newcommand{\GF}{{\rm G_F}}
\newcommand{\gPhi}{g_b^{\Phi}}
\newcommand{\mbMS}{\overline{m}_{b}}
\newcommand{\asrun}[1]{\as(#1)}
\newcommand{\NF}{N_F}
\newcommand{\gtPhi}{g_t^{\Phi}}
\newcommand{\Log}[1]{\log\left(#1\right)}
\newcommand{\smallaeff}{small $\alpha_{eff}$}
\newcommand{\ra}{\rightarrow}
\newcommand{\tautau}{\tau^+\tau^-}

\newcommand{\Quark}[4]{\ArrowLine(#1,#2)(#3,#4)}
\newcommand{\Gluino}[5]{\Gluon(#1,#2)(#3,#4){3}{#5}\Line(#1,#2)(#3,#4)}
\newcommand{\CrossedCircle}[6]{\BCirc(#1,#2){5}\Line(#3,#6)(#5,#4)\Line(#5,#6)(#3,#4)}
\newcommand{\Higgsino}[5]{\Photon(#1,#2)(#3,#4){3}{#5}\Line(#1,#2)(#3,#4)}
\newcommand{\myGluon}[5]{\Gluon(#1,#2)(#3,#4){3}{#5}}
\newcommand{\Squark}[4]{\DashArrowLine(#1,#2)(#3,#4){2}}
\newcommand{\Squarknoarrow}[4]{\DashLine(#1,#2)(#3,#4){2}}
\newcommand{\Quarknoarrow}[4]{\ArrowLine(#1,#2)(#3,#4)}

\newcommand{\trex}[1]{\textcolor{red}{#1}}

%%%%%%%%%%%%%%%%%%%%  Title  %%%%%%%%%%%%%%%%%%%%%%%%%%%%%%%%%%%%%%%%%

\vspace*{-2.5cm}

\begin{flushright}
KA--TP--18--2022 \\
PSI--PR--22--20 \\
\end{flushright}

\begin{center}
{\large \sc Top-Yukawa-induced Corrections to Higgs Pair Production}
\end{center}

\begin{center} {\sc Margarete M\"uhlleitner$^1$, Johannes Schlenk$^2$
and Michael Spira$^2$} \\[0.8cm]

\begin{small}
{\it \small
$^1$ Institut f\"ur Theoretische Physik, KIT, D--76128 Karlsruhe, Germany \\
$^2$ Paul Scherrer Institut, CH--5232 Villigen PSI, Switzerland}
\end{small}
\end{center}

%%%%%%%%%%%%%%%%%%%%  Abstract  %%%%%%%%%%%%%%%%%%%%%%%%%%%%%%%%%%%%%%%%%

\begin{abstract}
\noindent
Higgs-boson pair production at hadron colliders is dominantly mediated
by the loop-induced gluon-fusion process $gg\to HH$ that is
generated by heavy top loops within the Standard Model with a
minor per-cent level contamination of bottom-loop contributions. The QCD
corrections turn out to be large for this process. In this note, we
derive the top-Yukawa-induced part of the electroweak corrections to
this process and discuss their relation to an effective trilinear Higgs
coupling with integrated out top-quark contributions.
\end{abstract}

%%%%%%%%%%%%%%%%%%%%  Introduction  %%%%%%%%%%%%%%%%%%%%%%%%%%%%%%%%%%%%%%%%%

\section{Introduction}
%        ============
The discovery of a bosonic particle with a mass of $(125.09 \pm 0.24)$
GeV \cite{4} turned out to be in agreement with the Standard-Model (SM)
Higgs boson within the present uncertainties of all production and decay
modes. Its coupling strengths to SM gauge bosons, i.e.~$ZZ, W^+W^-$, and
fermion pairs as $\tau, \mu$ leptons and bottom quarks as well as the
loop-induced couplings to gluon and photon pairs, have been measured
with accuracies of $10-50\%$. All measurements are in agreement with the
SM predictions within their uncertainties \cite{5}.  In addition, there
are very strong indications that the newly discovered boson carries zero
spin and positive CP-parity, i.e.~possible deviations from these
hypotheses are strongly constrained by the accuracy of present
experimental data. Thus, there is increasing evidence that this particle
is indeed the long-sought SM Higgs boson.  Its discovery is of vital
importance for the consistency of the SM and the success of
the predictions for the precision electroweak observables which are in
striking agreement with measurements at LEP and SLC \cite{6}. The
discovery of a SM-like Higgs boson at the LHC completed the SM of
electroweak and strong interactions. The existence of the Higgs boson is
inherently related to the mechanism of spontaneous symmetry breaking
while preserving the full gauge symmetry and the renormalizability of
the SM \cite{7}, since the Higgs boson permits the SM particles to be
weakly interacting up to high-energy scales \cite{8}.
However, with the knowledge of the Higgs-boson mass all its properties
within the SM are uniquely fixed, i.e.~the SM does not allow the Higgs
couplings to the SM particles to deviate from their unique predictions.

The minimal model as realized in the SM requires the introduction of one
isospin doublet of Higgs fields that leads after spontaneous symmetry
breaking to the existence of one scalar Higgs boson.  A crucial
experimental goal is the measurement of the Higgs potential, since the
formation of a non-trivial ground state with a finite vacuum expectation
value of the Higgs field causes electroweak symmetry breaking so that
the experimental verification of the Higgs potential itself is of
highest interest. The parameters describing the Higgs potential are the
Higgs mass and self-interactions of the Higgs field.  The production of
Higgs-boson pairs is the first class of processes that offers the direct
access to the trilinear self-coupling of the Higgs boson as a first step
towards the reconstruction of the full Higgs potential. At
the Large Hadron Collider (LHC), the dominant Higgs-boson pair
production mechanism is provided by the gluon-fusion process $gg\to HH$,
while the other production modes as vector-boson fusion (VBF) $qq\to
qqHH$, double Higgs-strahlung $q\bar q\to W/Z + HH$ and double Higgs
bremsstrahlung off top quarks $q\bar q, gg\to t\bar tHH$ are suppressed
by at least one order of magnitude \cite{hhrev}. The individual
production cross sections roughly follow the pattern of
single-Higgs boson production but are in general smaller by about
three orders of magnitude. Since the trilinear Higgs coupling
contributes only to a subset of diagrams of each production process the
sensitivity to the trilinear Higgs coupling is reduced due to the
dominance of the continuum diagrams. The slope of the gluon-fusion cross
section as a function of the trilinear Higgs coupling $\lambda$ follows
the rough behaviour $\Delta\sigma/\sigma \sim - \Delta\lambda/\lambda$
around the SM prediction \cite{hhrev,nlohtl,hhpro}. This implies that the
uncertainties of the production cross section are immediately translated
to the uncertainties of the extracted trilinear self-coupling so that
the reduction of the theoretical uncertainties of the Higgs pair
production cross section is crucial for an accurate extraction of the
trilinear self-interaction from the experimental measurements. This
feature translates to a similar situation for the distributions as well.
The trilinear coupling develops a significant contribution for
Higgs-pair production closer to the production threshold, while it dies
out for large invariant Higgs-pair masses. In the last range, however,
statistics will be small in experiment so that the bulk of reconstructed
events will emerge from the region closer to the threshold.

The gluon-fusion mechanism $gg\to HH$ is mediated by top- and to a much
lesser extent bottom-quark loops, see Fig.~\ref{fg:hhdialo}. The full
next-to-leading-order (NLO) QCD corrections have been calculated by a
time-consuming numerical integration of the corresponding two-loop
integrals, since there are no systematic analytical methods to calculate
the corresponding two-loop integrals \cite{52,52a}. Similar to the
single-Higgs case they enhance the cross section by about 100\%. Because
the invariant mass of the final-state Higgs-boson pair is significantly
larger than in the single-Higgs case, the heavy top-quark limit (HTL)
works less reliably for Higgs-boson pairs. The full NLO QCD corrections
result in a decrease of the total cross section by about 15\%, due to
finite NLO top mass effects beyond the heavy-top limit, at the LHC for a
c.m.~energy of 14 TeV. This shows that the heavy-top limit for the
relative QCD corrections \cite{nlohtl} works still quite well for the
total cross section also in the Higgs-pair case. For the exclusive
cross section at large invariant
Higgs-pair masses, however, the finite mass effects at NLO can reach a
level of $-30\%$. The next-to-NLO (NNLO) QCD corrections to the total
cross section have been obtained in the heavy top-quark limit. They
imply an additional moderate rise of the total cross section by about
20\% \cite{53}. Recently, the next-to-NNLO (N$^3$LO) QCD corrections to
the total cross section became available and turned out to be small,
affecting the total cross section at the few per-cent level only
\cite{gghhn3lo}. NNLO top mass effects have been estimated to about 5\%
by means of a heavy top-quark expansion of the 2-loop virtual
corrections \cite{54}. Beyond NNLO, the
next-to-next-to-leading-logarithmic (NNLL) soft and collinear gluon
resummation contributes 5--10\% to the total cross section \cite{55}.
The factorization and renormalization scale dependence has been reduced
to about 5\%. In order to obtain an estimate of the residual theoretical
uncertainties, however, the uncertainties due to the scheme and scale
choice of the virtual top mass have to be taken into account as well.
These latter effects increase the theoretical uncertainties to a level
of 20--25\% \cite{52a}. The electroweak corrections to this process are
unknown.  They are expected in the 10\%-range for the total cross
section, but larger in the tails of the distributions.

In this work we investigate the electroweak corrections induced by the
top-Yukawa coupling as a uniquely defined contribution to the full
electroweak corrections. In Section \ref{sc:lo}, we will define our
notation and the corresponding leading-order (LO) result for $gg\to HH$.
In Section \ref{sc:eff}, we describe the effective Higgs (pair)
couplings to gluons in the HTL and the effective trilinear Higgs
coupling within the effective-potential approach, where the top
contributions are integrated out. Section \ref{sc:nlo} describes the NLO
calculation and Section \ref{sc:results} our results with a discussion
of our findings. In Section \ref{sc:conclusions}, we conclude.

\section{Higgs-boson pair production at leading order} \label{sc:lo}
%        ============================================
\begin{figure}[hbtp]
\begin{center}
\setlength{\unitlength}{1pt}
\begin{picture}(100,90)(100,0)
\Gluon(0,80)(50,80){3}{5}
\Gluon(0,20)(50,20){3}{5}
\ArrowLine(50,20)(50,80)
\ArrowLine(50,80)(100,80)
\ArrowLine(100,80)(100,20)
\ArrowLine(100,20)(50,20)
\DashLine(100,80)(150,80){5}
\DashLine(100,20)(150,20){5}
\put(155,76){$H$}
\put(155,16){$H$}
\put(-15,78){$g$}
\put(-15,18){$g$}
\put(30,48){$t,b$}
\end{picture}
\begin{picture}(100,90)(-20,0)
\Gluon(0,80)(50,80){3}{5}
\Gluon(0,20)(50,20){3}{5}
\ArrowLine(50,20)(50,80)
\ArrowLine(50,80)(100,50)
\ArrowLine(100,50)(50,20)
\DashLine(100,50)(150,50){5}
\DashLine(150,50)(200,80){5}
\DashLine(150,50)(200,20){5}
\put(205,76){$H$}
\put(205,16){$H$}
\put(-15,78){$g$}
\put(-15,18){$g$}
\put(30,48){$t,b$}
\put(146,46){\Large \textcolor{red}{$\bullet$}}
\put(146,60){\textcolor{red}{$\lambda$}}
\end{picture}
\setlength{\unitlength}{1pt}
\caption{\label{fg:hhdialo} \it Diagrams contributing to Higgs-boson
pair production via gluon fusion. The contribution of the trilinear
Higgs coupling is marked in red.}
\end{center}
\end{figure}
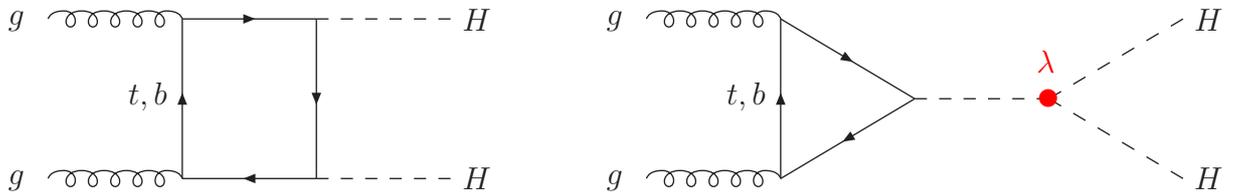
The LO Higgs pair production via gluon fusion is mediated by heavy
top-loop contributions and a marginal contribution of bottom loops, see
Fig.~\ref{fg:hhdialo}. In this work we neglect the bottom-loop
contributions and take into account the top loops only. The Higgs-boson
pair production cross section at LO is given by
\begin{equation}
\sigma_{\mathrm{LO}} = \int_{\tau_0}^1 d\tau~\frac{d{\cal
L}^{gg}}{d\tau}~\hat\sigma_{\mathrm{LO}}(Q^2 = \tau s) \,,
\end{equation}
where ${\cal L}^{gg}$ denotes the gluonic parton luminosity given in
terms of the gluon densities
$g(x,\mu_F)$,
\begin{equation}
\frac{d{\cal L}^{gg}}{d\tau} = \int_\tau^1 \frac{dx}{x} g(x,\mu_F)
g\left(\frac{\tau}{x},\mu_F\right)
\label{eq:lgg}
\end{equation}
at the factorization scale $\mu_F$, and the integration boundary is given
by $\tau_0=4M_H^2/s$, where $s$ denotes the hadronic center-of-mass
(c.m.) energy squared and $M_H$ the Higgs mass. The scale $Q^2=M_{HH}^2$
is defined in terms of the invariant mass $M_{HH}$ of the Higgs pair at
LO. The LO partonic cross section can be cast into the form
\begin{equation}
\hat\sigma_{LO} = \frac{G_F^2\alpha_s^2(\mu_R)}{512 (2\pi)^3}
\int_{\hat t_-}^{\hat t_+} d\hat t \Big[ | C_\triangle F_\triangle +
F_\Box|^2 + |G_\Box|^2 \Big] \,,
\label{eq:gghhlo}
\end{equation}
where the integration boundaries are given by
\begin{equation}
\hat t_\pm = -\frac{1}{2} \left[ Q^2 - 2M_H^2 \mp Q^2
\sqrt{1-4\frac{M_H^2}{Q^2}} \right] \, ,
\label{eq:tbound}
\end{equation}
and the symmetry factor 1/2 for the identical Higgs bosons in the
final state is included. The coefficient $C_\triangle = \lambda_{HHH}
v/(Q^2-M_H^2)$ involves the trilinear Higgs coupling that is related to
the Higgs mass and the vacuum expectation value (vev) $v$ at LO,
\begin{equation}
\lambda_{HHH} = 3\frac{M_H^2}{v} \,,
\label{eq:lambdaLO}
\end{equation}
where the vev is related to the Fermi constant $G_F = 1/(\sqrt{2}
v^2)$. The factor $\alpha_s(\mu_R)$ denotes the strong coupling at the
renormalization scale $\mu_R$.  The form factors $F_\triangle$ of the LO
triangle diagrams and $F_\Box, G_\Box$ of the LO box diagrams can be
found in Refs.~\cite{gghhlo}. In the HTL they approach simple
expressions, $F_\triangle \to 2/3$, $F_\Box \to -2/3$ and $G_\Box\to 0$.

\section{Effective Lagrangians} \label{sc:eff}
%        =====================
In this section we address the effective gluonic single- and
double-Higgs couplings as well as the effective Higgs self-couplings
after integrating out the heavy-top contributions, i.e.~the effective
couplings valid in the HTL at the leading order of an inverse large
top-mass expansion.

\subsection{Gluonic Higgs couplings}
%           =======================
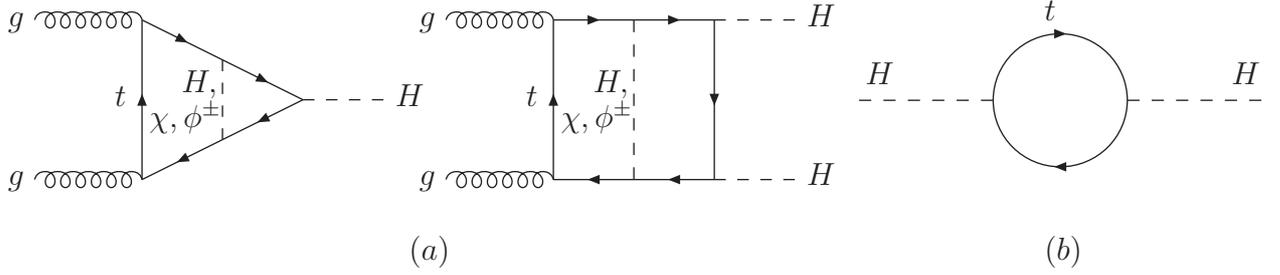
\begin{figure}[hbt]
\begin{center}
\begin{picture}(100,100)(70,0)
\Gluon(0,80)(40,80){3}{6}
\Gluon(0,20)(40,20){3}{6}
\ArrowLine(40,80)(70,65)
\ArrowLine(70,65)(100,50)
\ArrowLine(100,50)(70,35)
\ArrowLine(70,35)(40,20)
\ArrowLine(40,20)(40,80)
\DashLine(70,65)(70,35){5}
\DashLine(100,50)(130,50){5}
\put(135,48){$H$}
\put(55,53){$H,$}
\put(43,40){$\chi,\phi^\pm$}
\put(-10,18){$g$}
\put(-10,78){$g$}
\put(30,48){$t$}
\put(140,-10){$(a)$}
\end{picture}
\begin{picture}(100,100)(20,0)
\Gluon(0,80)(40,80){3}{6}
\Gluon(0,20)(40,20){3}{6}
\ArrowLine(40,80)(70,80)
\ArrowLine(70,80)(100,80)
\ArrowLine(100,80)(100,20)
\ArrowLine(100,20)(70,20)
\ArrowLine(70,20)(40,20)
\ArrowLine(40,20)(40,80)
\DashLine(70,80)(70,20){5}
\DashLine(100,80)(130,80){5}
\DashLine(100,20)(130,20){5}
\put(135,78){$H$}
\put(135,18){$H$}
\put(55,53){$H,$}
\put(43,40){$\chi,\phi^\pm$}
\put(-10,18){$g$}
\put(-10,78){$g$}
\put(30,48){$t$}
\end{picture}
\begin{picture}(100,100)(120,-50)
\DashLine(150,0)(200,0){5}
\ArrowArcn(225,0)(25,360,180)
%\ArrowArcn(200,0)(50,270,180)
\ArrowArcn(225,0)(25,540,360)
%\ArrowArcn(200,0)(50,450,360)
\DashLine(250,0)(300,0){5}
%\DashLine(200,50)(200,-50){5}
\put(153,6){$H$}
\put(220,30){$t$}
%\put(205,-2){$H,\chi,\phi^\pm$}
\put(290,6){$H$}
\put(220,-60){$(b)$}
\end{picture} \\
\end{center}
\caption[]{\it \label{fg:leff} Typical diagrams contributing to the
top-Yukawa-induced electroweak corrections to the effective Lagrangian: (a)
vertex corrections, (b) wave-function corrections. The fields $\chi,
\phi^\pm$ denote the pseudoscalar and charged would-be Goldstones.}
\end{figure}
In the HTL, the top-Yukawa-induced electroweak corrections to the
effective $Hgg$ and $HHgg$ couplings can be obtained as
\begin{equation}
{\cal L}_{eff} = C_1 \frac{\alpha_s}{12\pi} G^{a\mu\nu} G^a_{\mu\nu}
\log\left( 1+C_2 \frac{H}{v}\right) \,,
\end{equation}
where $G^a_{\mu\nu}$ denotes the gluonic field-strength tensor and $H$
the SM Higgs field. The radiatively corrected coefficients are given by
\begin{eqnarray}
C_1 & = & 1 - 3 x_t  + {\cal O}(x_t^2) \nonumber \\
C_2 & = & 1 + \frac{7}{2} x_t + {\cal O}(x_t^2) \,,
\end{eqnarray}
with $x_t = G_F m_t^2/(8\sqrt{2}\pi^2)$, where $C_1$ describes the
genuine corrections to the $Hgg$ and $HHgg$ vertices \cite{gghelw} (see
Fig.~\ref{fg:leff}a) and $C_2$ the universal top-Yukawa-induced
correction related to the Higgs wave-function and vacuum expectation
value \cite{huniv} (see Fig.~\ref{fg:leff}b). This yields the explicit
effective $Hgg$ and $HHgg$ couplings,
\begin{equation}
{\cal L}_{eff} = \frac{\alpha_s}{12\pi} G^{a\mu\nu} G^a_{\mu\nu} \left\{
(1+\delta_1) \frac{H}{v} + ( 1 + \eta_1) \frac{H^2}{2v^2} + {\cal
O}(H^3) \right\}
\label{eq:leff}
\end{equation}
where
\begin{eqnarray}
\delta_1 & = & \frac{x_t}{2} + {\cal O}(x_t^2) \,, \qquad\qquad
\eta_1   = 4 x_t + {\cal O}(x_t^2) \,.
\end{eqnarray}
This effective Lagrangian describes the electroweak corrections induced
by $x_t$ to the $Hgg$ and $HHgg$ vertices in the HTL and will be used in
this limit in the following. We would like to point out explicitly that
the square root of the wave-function counterterm of the external Higgs
boson(s) is already taken into account in this effective Lagrangian.

\subsection{Higgs self-couplings}
%           ====================
The starting point of effective Higgs self-couplings is the effective
one-loop corrected Higgs potential involving virtual top-quark
effects of the SM \cite{selfeff},
\begin{eqnarray}
V_{eff} & = & V_0 + V_1 \nonumber \\
V_0 & = & \mu_0^2 |\phi|^2 + \frac{\lambda_0}{2} |\phi|^4
\nonumber \\
V_1 & = & \frac{3\overline{m}_t^4}{16\pi^2} C_\epsilon \left(
\frac{1}{\epsilon} + \log \frac{\bar\mu^2}{\overline{m}_t^2} + \frac{3}{2}
\right) \,,
\end{eqnarray}
with the bare Higgs self-coupling $\lambda_0$, the SM Higgs doublet in
unitary gauge,
\begin{equation}
\phi = \frac{1}{\sqrt{2}} \left(\begin{array}{c} 0 \\ v+H
\end{array} \right)
\end{equation}
the loop coefficient
\begin{equation}
C_\epsilon = \Gamma(1+\epsilon) (4\pi^2)^\epsilon
\end{equation}
and the field-dependent top-mass parameter
\begin{equation}
\overline{m}_t = m_t \left( 1+\frac{H}{v}\right) \,.
\end{equation}
The expression above for the effective Higgs potential involves the 't
Hooft scale $\bar\mu$. After minimization of the effective Higgs potential
by the tadpole equation,
\begin{eqnarray}
\mu_0^2 & = & -\frac{\lambda_0}{2} v^2 + \delta\mu^2 \nonumber \\
\delta\mu^2 & = & -\frac{3m_t^4}{4\pi^2 v^2} C_\epsilon \left\{
\frac{1}{\epsilon} + \log \frac{\bar\mu^2}{m_t^2} + 1\right\}
\end{eqnarray}
and the renormalization of the Higgs mass,
\begin{eqnarray}
M_{H0}^2 & = & \lambda_0 v^2 = \lambda v^2 +
(\delta\lambda) v^2 = \lambda v^2 + \delta M_H^2 \nonumber \\
\delta M_H^2 & = & -\frac{3m_t^4}{2\pi^2 v^2} C_\epsilon \left\{
\frac{1}{\epsilon} + \log \frac{\bar\mu^2}{m_t^2} \right\} \,,
\end{eqnarray}
the effective Higgs trilinear (quartic) self-coupling can be obtained by
the third (fourth) derivative of this effective Higgs potential with
respect to the physical Higgs field $H$,
\begin{eqnarray}
\lambda_{HHH}^{eff} & = & 3 \frac{M_H^2}{v} + \Delta\lambda_{HHH} \,,
\qquad\qquad
\lambda_{HHHH}^{eff} = 3 \frac{M_H^2}{v^2} + \Delta\lambda_{HHHH} \,,
\label{eq:lameff}
\end{eqnarray}
with
\begin{eqnarray}
\Delta\lambda_{HHH} & = & - \frac{3 m_t^4}{\pi^2 v^3} \,, \qquad\qquad
\Delta\lambda_{HHHH} = - \frac{12 m_t^4}{\pi^2 v^4} \,.
\label{eq:dlam}
\end{eqnarray}
These effective NLO couplings are the relevant Higgs
self-interactions in the HTL and will be compared with the full
triple-vertex corrections within this work.

\section{Top-Yukawa-induced electroweak corrections to Higgs pair production}
%        ===================================================================
\label{sc:nlo}
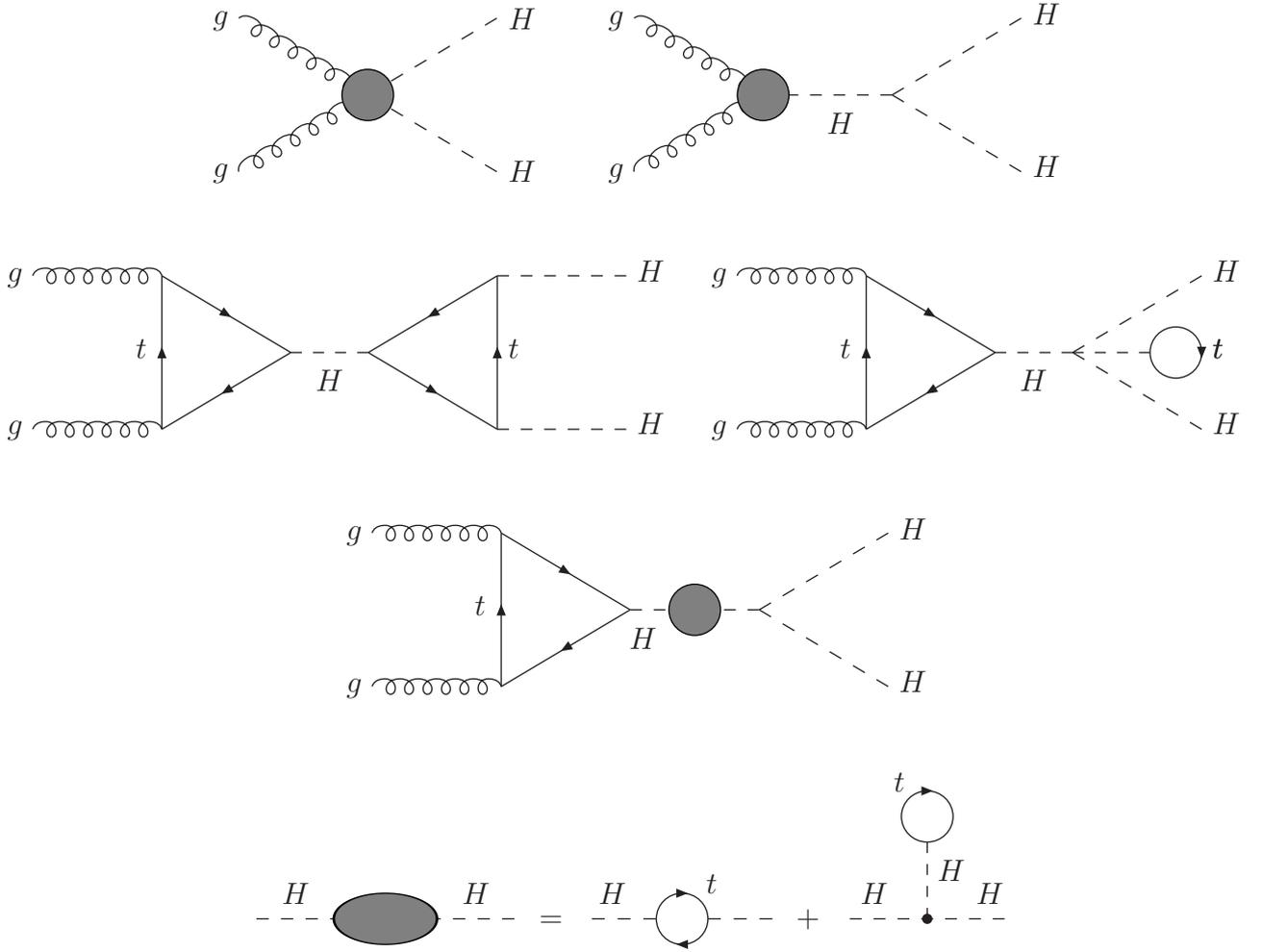
\begin{figure}[hbt]
\begin{center}
\begin{picture}(100,50)(50,30)
\Gluon(0,80)(50,50){3}{6}
\Gluon(0,20)(50,50){3}{6}
\DashLine(50,50)(100,80){5}
\DashLine(50,50)(100,20){5}
\GCirc(50,50){10}{0.5}
\put(105,76){$H$}
\put(105,16){$H$}
\put(-10,18){$g$}
\put(-10,78){$g$}
\end{picture}
\begin{picture}(100,50)(0,30)
\Gluon(0,80)(50,50){3}{6}
\Gluon(0,20)(50,50){3}{6}
\DashLine(50,50)(100,50){5}
\DashLine(100,50)(150,80){5}
\DashLine(100,50)(150,20){5}
\GCirc(50,50){10}{0.5}
\put(75,35){$H$}
\put(155,78){$H$}
\put(155,18){$H$}
\put(-10,18){$g$}
\put(-10,78){$g$}
\end{picture} \\
\begin{picture}(100,100)(130,30)
\Gluon(0,80)(50,80){3}{6}
\Gluon(0,20)(50,20){3}{6}
\ArrowLine(50,80)(100,50)
\ArrowLine(100,50)(50,20)
\ArrowLine(50,20)(50,80)
\DashLine(100,50)(130,50){5}
\ArrowLine(180,80)(130,50)
\ArrowLine(130,50)(180,20)
\ArrowLine(180,20)(180,80)
\DashLine(180,80)(230,80){5}
\DashLine(180,20)(230,20){5}
\put(110,35){$H$}
\put(235,78){$H$}
\put(235,18){$H$}
\put(-10,18){$g$}
\put(-10,78){$g$}
\put(40,48){$t$}
\put(185,48){$t$}
\end{picture}
\begin{picture}(100,100)(-40,30)
\Gluon(0,80)(50,80){3}{6}
\Gluon(0,20)(50,20){3}{6}
\ArrowLine(50,80)(100,50)
\ArrowLine(100,50)(50,20)
\ArrowLine(50,20)(50,80)
\DashLine(100,50)(130,50){5}
\DashLine(130,50)(180,80){5}
\DashLine(130,50)(160,50){5}
\DashLine(130,50)(180,20){5}
\ArrowArcn(170,50)(10,540,180)
\put(110,35){$H$}
\put(185,78){$H$}
\put(185,18){$H$}
\put(185,48){$t$}
\put(40,48){$t$}
\put(185,48){$t$}
\put(-10,18){$g$}
\put(-10,78){$g$}
\end{picture} \\
\begin{picture}(100,100)(50,30)
\Gluon(0,80)(50,80){3}{6}
\Gluon(0,20)(50,20){3}{6}
\ArrowLine(50,80)(100,50)
\ArrowLine(100,50)(50,20)
\ArrowLine(50,20)(50,80)
\DashLine(100,50)(150,50){5}
\GOval(125,50)(10,10)(0){0.5}
\DashLine(150,50)(200,80){5}
\DashLine(150,50)(200,20){5}
\put(100,35){$H$}
\put(205,78){$H$}
\put(205,18){$H$}
\put(40,48){$t$}
\put(-10,18){$g$}
\put(-10,78){$g$}
\end{picture} \\
\begin{picture}(130,110)(80,-10)
\DashLine(0,0)(100,0){5}
\GOval(50,0)(10,20)(0){0.5}
\DashLine(130,0)(155,0){5}
\ArrowArcn(165,0)(10,360,180)
\ArrowArcn(165,0)(10,540,360)
\DashLine(175,0)(200,0){5}
\DashLine(230,0)(290,0){5}
\DashLine(260,0)(260,30){5}
\ArrowArcn(260,40)(10,270,-90)
\Vertex(260,0){2}
\put(110,-3){$=$}
\put(210,-3){$+$}
\put(10,6){$H$}
\put(80,6){$H$}
\put(133,6){$H$}
\put(175,10){$t$}
\put(235,6){$H$}
\put(280,6){$H$}
\put(265,15){$H$}
\put(248,50){$t$}
\end{picture} \\
\end{center}
\caption[]{\it \label{fg:elwcorr} Generic diagrams describing the
top-Yukawa-induced electroweak corrections to Higgs-boson pair production via
gluon fusion. The blobs of the first two diagrams are determined
by the effective Lagrangian of Eq.~(\ref{eq:leff}) in the HTL.}
\end{figure}
The top-Yukawa-induced electroweak corrections arise from NLO diagrams
involving top-quark loops as shown in Fig.~\ref{fg:elwcorr}, where the
tadpole diagrams are displayed explicitly. For simplicity, we will use
the relative corrections of Eq.~(\ref{eq:leff}) to the $ggH$ and
$ggHH$ vertices
in the HTL, while the radiative corrections to the triple-Higgs vertex
and Higgs self-energies are treated with full top-mass dependence. Since
there are no real corrections at the NLO electroweak level, the
radiative corrections can be implemented by a shift of the LO form
factors of Eq.~(\ref{eq:gghhlo}),
\begin{eqnarray}
C_\triangle F_\triangle & \to & C_\triangle F_\triangle
(1+\Delta_\triangle) \nonumber \\
F_\Box      & \to & F_\Box      (1+\Delta_\Box) \,,
\label{eq:signlo}
\end{eqnarray}
while the LO form factor $G_\Box$ does not receive
top-Yukawa-induced electroweak corrections in our approach,
since $G_\Box$ vanishes in the HTL. The top-Yukawa-induced radiative
corrections in Eq.~(\ref{eq:signlo}) read as
\begin{eqnarray}
\Delta_\triangle & = & \delta_1 + \Delta_{HHH} \nonumber \\
\Delta_\Box      & = & \eta_1 \,,
\end{eqnarray}
where the vertex, self-energy and counterterm corrections are given by
\begin{eqnarray}
\Delta_{HHH} & = & \Delta_{vertex} + \Delta_{self} + \Delta_{CT}
\nonumber \\
\Delta_{vertex} & = & \frac{m_t^4}{v^2M_H^2} \frac{8}{(4\pi)^2} \left\{
B_0(Q^2;m_t,m_t) + 2 B_0(M_H^2;m_t,m_t) \right. \nonumber \\
& &  \left. + \left( 4m_t^2-\frac{Q^2+2M_H^2}{2}\right)
C_0(Q^2,M_H^2,M_H^2;m_t,m_t,m_t) \right\} + \frac{T_1}{vM_H^2} \nonumber \\
\Delta_{self} & = & \frac{\Sigma_H(Q^2)}{Q^2-M_H^2}
+\frac{1}{2} \Sigma_{H}'(M_H^2) \nonumber \\
\Delta_{CT} & = & \frac{\delta M_H^2}{Q^2-M_H^2} + \frac{\delta
\lambda_{HHH}}{\lambda_{HHH}} \,.
\label{eq:dhhh}
\end{eqnarray}
We are adopting the scalar integrals in $n=4-2\epsilon$ dimensions,
\begin{eqnarray}
A_0(m) & = & \frac{(4\pi)^2}{i}~\bar\mu^{4-n} \int
\frac{d^nk}{(2\pi)^n}
\frac{1}{k^2-m^2} \nonumber \\
B_0(p^2;m_1,m_2) & = & \frac{(4\pi)^2}{i}~\bar\mu^{4-n} \int
\frac{d^nk}{(2\pi)^n} \frac{1}{(k^2-m_1^2) [(k+p)^2-m_2^2]} \nonumber \\
B_0'(p^2;m_1,m_2) & = & \frac{\partial}{\partial p^2} B_0(p^2;m_1,m_2)
\nonumber \\
C_0(p_1^2,p_2^2,(p_1+p_2)^2;m_1,m_2,m_3) & = &
\frac{(4\pi)^2}{i}~\bar\mu^{4-n} \int \frac{d^nk}{(2\pi)^n}
\frac{1}{(k^2-m_1^2) [(k+p_1)^2-m_2^2]} \nonumber \\
& & \hspace*{3.5cm} \times \frac{1}{[(k+p_1+p_2)^2-m_3^2]} \,.
\end{eqnarray}
In the expression of Eq.~(\ref{eq:dhhh}), the self-energy
$\Sigma_H(Q^2)$ and its derivative $\Sigma'_H(Q^2)$, the tadpole
term $T_1/v$, the trilinear Higgs-coupling counterterm
$\delta\lambda_{HHH}$ and the Higgs-mass counterterm $\delta M_H^2$ are
given by
\begin{eqnarray}
\Sigma_H(Q^2) & = & 3\,\frac{T_1}{v} + 6\,\frac{m_t^2}{(4\pi)^2 v^2} \left\{
2A_0(m_t) + (4m_t^2-Q^2) B_0(Q^2; m_t, m_t)\right\} + {\cal O}(m_t^0)
\nonumber \\
\Sigma'_H(Q^2) & = & 6\,\frac{m_t^2}{(4\pi)^2 v^2} \left\{ (4m_t^2-Q^2)
B'_0(Q^2; m_t, m_t) - B_0(Q^2; m_t, m_t) \right\} + {\cal O}(m_t^0)
\nonumber \\
\frac{T_1}{v} & = & -12\,\frac{m_t^2}{(4\pi)^2 v^2} A_0(m_t) \nonumber \\
\frac{\delta\lambda_{HHH}}{\lambda_{HHH}} & = & \frac{\delta
M_H^2}{M_H^2} + \frac{1}{2} \frac{\Sigma_W(0)}{M_W^2} \nonumber \\
\frac{\Sigma_W(0)}{M_W^2} & = & 2\,\frac{T_1}{v M_H^2} +
\frac{2m_t^2}{(4\pi)^2 v^2} \left\{ B_0(0;m_t,0) + 2 B_0(0;m_t,m_t) +
m_t^2 B_0'(0;m_t,0) \right\} + {\cal O}(m_t^0) \nonumber \\
\delta M_H^2 & = & -\Sigma_H(M_H^2) \,,
\end{eqnarray}
where the self-energies $\Sigma_H, \Sigma_W$ and the Higgs-mass
counterterm include tadpole contributions as well, and we only kept
terms of ${\cal O}(m_t^4)$ and ${\cal O}(m_t^2)$ for the
counterterms to be consistent. For
the calculation, we have used the alternative tadpole-scheme of
Ref.~\cite{FJ}\footnote{We have checked explicitly that in the
conventional approach of using a tadpole counterterm to cancel all
tadpole diagrams, we arrive at the same result for $\Delta_{HHH}$ due to
the residual tadpole contribution to the counterterm for the trilinear
Higgs coupling $\lambda_{HHH}$ \cite{dendit}.} and implemented the electroweak
parameters in the $G_F$ scheme, i.e.~choosing $G_F, M_Z, M_W$ as input
parameters for the electroweak gauge sector, while the Weinberg angle
$\theta_W$ and the QED coupling $\alpha$ are derived quantities. In
addition, we have taken into account that the effective Lagrangian of
Eq.~(\ref{eq:leff}) contains the wave-function renormalization of the
external Higgs fields that has to be compensated in the corrections
$\Delta_{HHH}$ to avoid double counting. Within our
electroweak renormalization, the trilinear coupling is given by its LO
expression in terms of the renormalized Higgs mass and vacuum
expectation value of Eq.~(\ref{eq:lambdaLO}).
We will compare the explicit NLO result of Eq.~(\ref{eq:signlo}) to the
corresponding one using the effective trilinear coupling
$\lambda_{HHH}^{eff}$, i.e.~adding the corresponding matching term
\begin{eqnarray}
\Delta_{HHH} & \to & \Delta_{HHH} + \Delta_\lambda \nonumber \\
\Delta_\lambda & = & -\frac{\Delta\lambda_{HHH}}{\lambda_{HHH}} = 16
\frac{m_t^4}{(4\pi)^2 v^2 M_H^2}
\label{eq:matching}
\end{eqnarray}
with $\Delta\lambda_{HHH}$ of Eq.~(\ref{eq:dlam}) to avoid double
counting and using the
effective coupling $\lambda_{HHH} \to \lambda_{HHH}^{eff}$ of
Eq.~(\ref{eq:lameff}) for the triangle coefficient $C_\triangle$ in
Eq.~(\ref{eq:gghhlo}) in both the LO and NLO expressions.

The relative electroweak corrections to the Higgs-pair production cross
section are defined by expanding the expression of Eq.~(\ref{eq:gghhlo})
up to NLO by using the corrected form factors of Eq.~(\ref{eq:signlo})
at the parton level,
\begin{eqnarray}
\hat\sigma_{NLO} & = & \hat\sigma_{LO} + \Delta\hat\sigma \nonumber \\
\Delta\hat\sigma & = & \frac{G_F^2\alpha_s^2(\mu_R)}{512 (2\pi)^3}
\int_{\hat t_-}^{\hat t_+} d\hat t~2 \Re e~
\Big\{ (C_\triangle F_\triangle + F_\Box)^* (C_\triangle
F_\triangle\Delta_\triangle + F_\Box\Delta_\Box) \Big\}
\end{eqnarray}
such that the hadronic cross section is corrected as
\begin{eqnarray}
\sigma_{NLO} & = & \sigma_{LO}~(1+\delta_{elw}) \nonumber \\
\delta_{elw} & = & \frac{\Delta\sigma}{\sigma_{LO}} \nonumber \\
\Delta\sigma & = & \int_{\tau_0}^1 \frac{{\cal L}^{gg}}{d\tau}
\Delta\hat\sigma
\end{eqnarray}
Within this expression we will either use the LO expression of the
triple Higgs coupling $\lambda_{HHH}$ of Eq.~(\ref{eq:lambdaLO}) or the
radiatively-corrected effective coupling $\lambda^{eff}_{HHH}$ of
Eq.~(\ref{eq:lameff}) with the according form of the radiative
corrections as shown in Eq.~(\ref{eq:matching}).

\section{Results} \label{sc:results}
%        =======
\begin{figure}[hbt]
\begin{center}
\begin{picture}(150,245)(0,0)
\put(-150,-160.0){\includegraphics{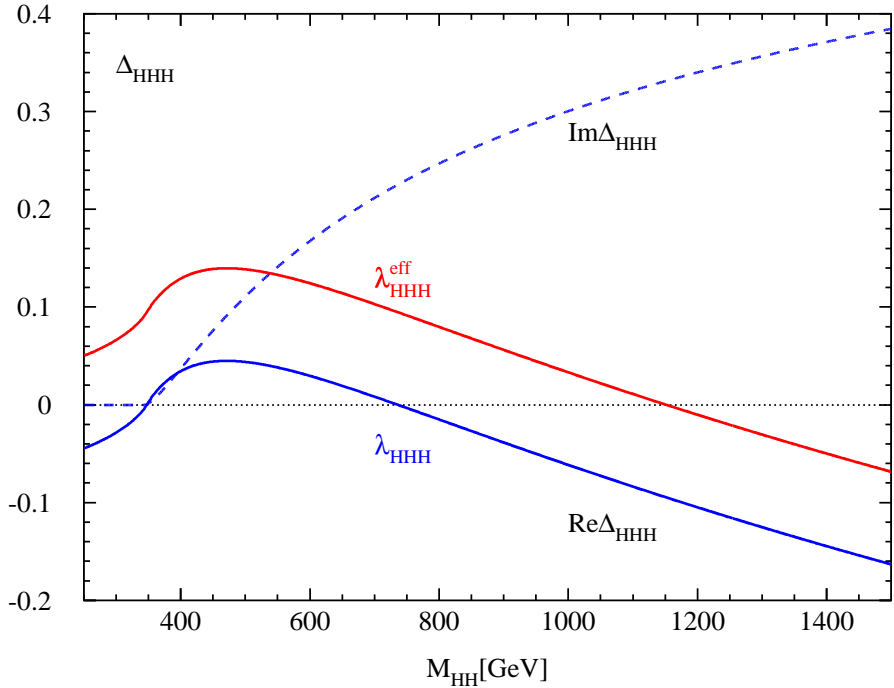}}
\end{picture}
\caption{\it The relative top-Yukawa-induced electroweak correction
factor $\Delta_{HHH}$ as a function of the invariant Higgs-pair mass
$M_{HH}$. The full blue curve shows the real part of $\Delta_{HH}$ and the
dashed blue on the imaginary part. The red curve exhibits the real part
after introducing the effective trilinear coupling $\lambda_{HHH}^{eff}$
of Eq.~(\ref{eq:lameff}) and adding the shift of Eq.~(\ref{eq:matching}).
To guide the eye the dotted black curve has been added as the zero-line.}
\label{fg:dhhh}
\end{center}
\end{figure}
\begin{figure}[hbt]
\begin{center}
\begin{picture}(150,245)(0,0)
\put(-150,-160.0){\includegraphics{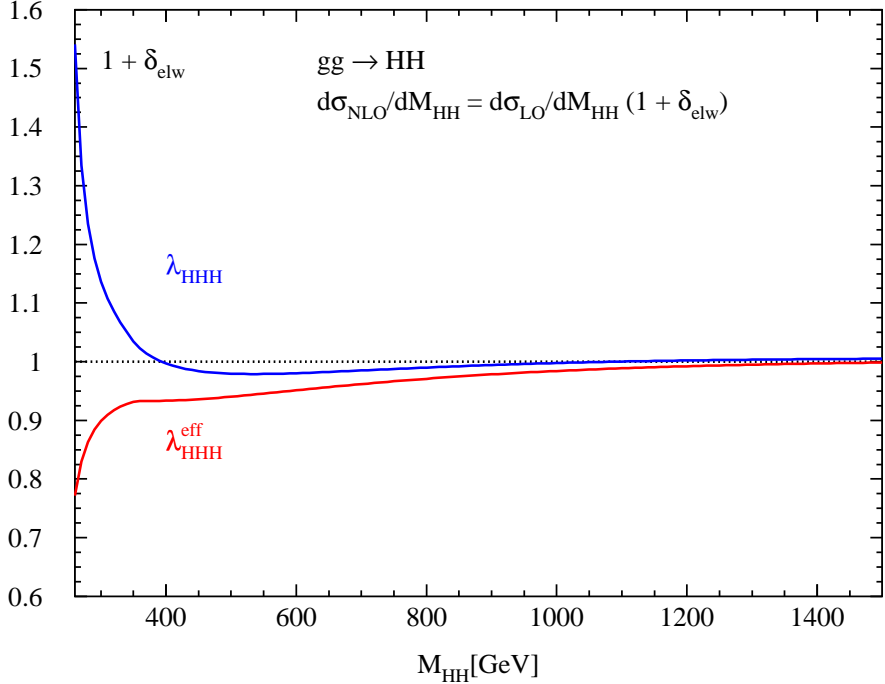}}
\end{picture}
\caption{\it The relative top-Yukawa-induced electroweak corrections to
the differential Higgs-pair production cross section as a function of
the invariant Higgs-pair mass $M_{HH}$. The blue curve shows the
electroweak correction factor using the LO trilinear Higgs coupling of
Eq.~(\ref{eq:lambdaLO}) and the red curve the corrections factor
involving the effective coupling $\lambda_{HHH}^{eff}$ of
Eq.~(\ref{eq:lameff}). The black dotted line at the value 1 is
inserted to guide the eye. The electroweak corrections factor is
independent of the hadronic c.m.~energy and scale choices in the QCD
part of the differential cross section $d\sigma/dM_{HH}$ so that it is
valid for any hadronic energy as a pure rescaling factor.}
\label{fg:corr}
\end{center}
\end{figure}
For our numerical analysis we work at a c.m.~energy of 14 TeV at the LHC
and use a top pole mass of $m_t=172.5$ GeV according to the conventions
of the LHC Higgs Working Group \cite{yr4}. The Fermi constant is chosen
as $G_F = 1.1663787 \times 10^{-5}~{\rm GeV}^{-2}$, the strong coupling
as $\alpha_s(M_Z) = 0.118$ and the Higgs mass as $M_H=125$ GeV. We are
using PDF4LHC15 parton densities.

The (complex) electroweak correction factor $\Delta_{HHH}$ of
Eq.~(\ref{eq:dhhh}) is shown in Fig.~\ref{fg:dhhh} as a function of the
invariant Higgs-pair mass $M_{HH}$. The full lines denote the real parts
and the dashed line the imaginary part. The blue curves exhibit the
real and imaginary parts of $\Delta_{HHH}$ in terms of the LO trilinear
Higgs coupling, while the red curve shows the correction factor
after introducing the effective coupling $\lambda_{HHH}^{eff}$. The size
of the correction factor shows that the effective trilinear coupling
does not capture the dominant part of the electroweak corrections so
that its use is not supported by our results.

The relative electroweak corrections originating from the
top-Yukawa-induced contributions are shown in Fig.~\ref{fg:corr} for the
differential cross section as a function of the invariant Higgs-pair
mass $M_{HH}$. The radiative corrections close to the production
threshold turn out to be large. This is due to the vanishing of the
matrix element in the leading term in the large inverse top-mass
expansion of the LO expression of Eq.~(\ref{eq:gghhlo}) so that the LO
matrix element is highly suppressed at threshold. This suppression,
however, is lifted by the radiative corrections to the effective
trilinear Higgs coupling $\lambda^{eff}_{HHH}$ or, equivalently, the
mismatch of electroweak corrections to the triangle and box diagrams.
However, Fig.~\ref{fg:corr} does not support the use of the effective
trilinear Higgs coupling $\lambda^{eff}_{HHH}$ to improve the
perturbative result.  Thus, the naive argument that the effective
trilinear Higgs coupling induces a SM contribution to $\kappa_\lambda$,
\begin{eqnarray}
\lambda^{eff}_{HHH} & = & \kappa_\lambda \lambda_{HHH} \nonumber \\
\lambda_{HHH} & = & 3~\frac{M_H^2}{v} \nonumber \\
\kappa_\lambda & = & 1 - \frac{m_t^4}{\pi^2 v^2 M_H^2} \approx 0.91
\end{eqnarray}
is not supported by our results, but the inclusion of the complete
electroweak corrections is mandatory instead. We observe that the
electroweak corrections appear
with opposite sign close to the threshold between the options of using the LO and the effective trilinear coupling.

The effect of the top-Yukawa-induced electroweak corrections on the
total integrated hadronic cross section amounts to
\begin{eqnarray}
\sigma & = & K_{elw} \times \sigma_{LO} \nonumber \\
K_{elw} & \approx & 1.002 \hspace*{1.2cm} \mbox{($\lambda_{HHH}$)} \nonumber \\
K^{eff}_{elw} & \approx & 0.938 \hspace*{1cm} \mbox{($\lambda^{eff}_{HHH}$)}
\end{eqnarray}
so that the corrections induce an effect of about 0.2\% on the total
cross section, if the LO-like trilinear Higgs coupling
$\lambda_{HHH}$ is adopted. The bulk of these corrections {\it cannot}
be absorbed in the effective triple Higgs coupling, but the latter
option leads to an artificial increase of the relative electroweak
corrections.

\section{Conclusions} \label{sc:conclusions}
%        ===========
In this note we have investigated the electroweak corrections to
Higgs-pair production via gluon fusion induced by top-quark
contributions. While keeping the full top-mass dependence in the
triple-Higgs vertex and self-energy corrections, we have worked in the
HTL for the radiative corrections to the effective $ggH(H)$ vertices for
the relative corrections. The top-Yukawa-induced NLO electroweak
corrections to the total gluon-fusion cross section amount to about
0.2\%. After integrating out the top-quark contributions an
effective trilinear Higgs coupling can be defined in terms of the
effective Higgs potential that is dressed with contributions scaling
with the fourth power of the top mass. This is known already starting
from the Coleman--Weinberg potential \cite{selfeff}. This effective
trilinear Higgs coupling can be introduced in the full calculation of
electroweak corrections as well and leads to a modification of the
counterterms in order to remove potential double counting of
corrections. However, introducing this effective coupling the remaining
electroweak corrections turn out to be larger than in the case of the
LO-like triple Higgs coupling.  \\

\noindent
{\bf Acknowledgments} \\
The authors are indebted to A.~Djouadi and P.~Gambino for very helpful
discussions. The research of M.M. is supported by the Deutsche
Forschungsgemeinschaft (DFG, German Research Foundation) under grant
396021762--TRR 257. The work of J.S.~is supported by the Swiss National
Science Foundation (SNSF).

%%%%%%%%%%%%%%   Bibliography   %%%%%%%%%%%%%%%%%%%%%%%%%%%%%%%%%%%%%%%%%

\end{document}